\documentclass[a4paper]{article}
\usepackage{graphicx}
\usepackage{twocolceurws}
\usepackage[numbers]{natbib}
\usepackage{titlesec}
\usepackage{enumitem}
\usepackage{booktabs}
\usepackage{float}
\usepackage{cleveref}
\usepackage{multirow}
\usepackage{todonotes}
\usepackage{tikz}
\usepackage{caption}
\usepackage{natbib}

\title{Explainability and AI Confidence in Clinical Decision Support Systems: Effects on Trust, Diagnostic Performance, and Cognitive Load in Breast Cancer Care}

\author{
Olya Rezaeian \\ Dept. of Systems and Enterprises\\
                Stevens Institute of Technology   
\and
Alparslan Emrah Bayrak \\ Dept. of Mechanical Engineering and Mechanics\\ Lehigh University  
\and
\\
Onur Asan \\ Dept. of Systems and Enterprises\\
                Stevens Institute of Technology    
}

\institution{}
\makeatletter
\def\copyrightspace{} 
\gdef\@thanks{} 
\makeatother
\begin{document}

\maketitle

\begin{abstract}
Artificial Intelligence (AI) has demonstrated potential in healthcare, particularly in enhancing diagnostic accuracy and decision-making through Clinical Decision Support Systems (CDSSs). However, the successful implementation of these systems relies on user trust and reliance, which can be influenced by explainable AI. This study explores the impact of varying explainability levels on clinicians’ trust, cognitive load, and diagnostic performance in breast cancer detection. Utilizing an interrupted time series design, we conducted a web-based experiment involving 28 healthcare professionals. The results revealed that high confidence scores substantially increased trust but also led to overreliance, reducing diagnostic accuracy. In contrast, low confidence scores decreased trust and agreement while increasing diagnosis duration, reflecting more cautious behavior. Some explainability features influenced cognitive load by increasing stress levels. Additionally, demographic factors such as age, gender, and professional role shaped participants' perceptions and interactions with the system. This study provides valuable insights into how explainability impact clinicians' behavior and decision-making. The findings highlight the importance of designing AI-driven CDSSs that balance transparency, usability, and cognitive demands to foster trust and improve integration into clinical workflows.
\end{abstract}

\section{Introduction}

In recent years, Artificial Intelligence (AI) has emerged as a transformative technology across various fields, offering the potential to significantly improve processes and outcomes. Healthcare has become a primary focus for AI research, given the vast amounts of data it generates and the opportunities to use this information to enhance patient care. AI systems have shown promise in improving diagnosis \cite{mckinney_international_2020,micocci_attitudes_2021, janowczyk_deep_2016, nahata_deep_2020}, treatment planning \cite{mcintosh_voxel-based_2016, choudhury_effect_2022}, and workflow efficiency, and also reducing human errors\cite{cartolovni_ethical_2022,gaube_as_2021,topol_high-performance_2019}. One area where AI is making significant progress is in breast cancer detection, with Clinical Decision Support Systems (CDSS) playing a key role. These AI-powered tools assist radiologists and oncologists by analyzing medical images, identifying abnormalities, and providing diagnostic recommendations\cite{mckinney_international_2020}. 

Despite the potential benefits of integrating AI into Clinical Decision Support Systems (CDSSs), their value ultimately depends on user acceptance and adoption \cite{asan_artificial_2020, parasuraman_humans_1997, ghazizadeh_extending_2012, davis_perceived_1989}. Trust plays a pivotal role in this process, directly shaping clinicians’ willingness to rely on these systems \cite{choudhury_effect_2022, glikson_human_2020}. As AI continues to advance and find applications in various healthcare domains, addressing trust-related barriers has become critical to ensuring the successful implementation of AI-based CDSSs \cite{asan_artificial_2020}. Among the factors influencing trust, explainability stands out as an important element \cite{tucci_factors_2022}. AI systems often operate as ''black boxes``, offering results without clear reasoning, which can lead to skepticism among clinicians. When AI systems provide explanations for their recommendations, they are more likely to be trusted and integrated into clinical workflows. 

On the other hand, the integration of explainability in AI-driven CDSSs requires an examination of their role in improving the user experience \cite{miller_effects_2009}. It is crucial to understand how explainability affects factors such as the mental demands and stress levels. 

Additionally, system confidence scores play a crucial role in shaping trust\cite{zhang_effect_2020, antifakos_towards_2005}. Confidence scores, which reflect the AI system’s level of certainty in its recommendations, serve as valuable information for clinicians to evaluate the reliability of the outputs. The varying levels of confidence scores, ranging from low to high certainty, can substantially impact how clinicians interpret and act upon AI recommendations.

To investigate the effects of varying levels of explainability on mental demand and stress, as well as the impact of system confidence scores on trust and performance, we designed a user-centered experiment. This study evaluated how clinicians interacted with our custom-built web application, which integrated an AI-based Clinical Decision Support System (CDSS).

The web application presented medical cases accompanied by AI-generated recommendations with varying levels of explainability and confidence scores. Participants included radiologists, oncologists, and other healthcare professionals who engaged with the system in a controlled setting. During the experiment, we collected data on trust, cognitive load, stress levels, and diagnostic performance through surveys and system interactions.

This study is guided by three key research questions: (1) How do different levels of explainability in AI-based Clinical Decision Support Systems affect clinicians' cognitive load? and (2) How do varying system confidence scores influence trust and diagnostic performance? (3) How do clinicians' demographic characteristics influence their mental demand and stress in AI-based clinical decision support systems (CDSS) in breast cancer detection?

By addressing these questions, we aim to deepen our understanding of how design features like explainability and confidence scores shape the interaction between clinicians and AI systems. Ultimately, this research seeks to inform the development of AI-based CDSS that are not only effective but also trusted and seamlessly integrated into clinical workflows.

\section{Related Work}

\subsection{Technology Adoption And Trust}

Numerous studies in the literature have explored various aspects of technology adoption, leading to the development of models such as the Technology Acceptance Model (TAM) \cite{davis_perceived_1989}, widely applied in healthcare studies \cite{jalali_sepehr_exploring_2024}, Task-Technology fit (TTF) \cite{goodhue_task-technology_1995} and the Unified Theory of Acceptance and Use of Technology (UTAUT) \cite{venkatesh_user_2003}. TAM, in particular, offers a foundational framework for understanding users' adoption of technology by identifying two primary constructs, Perceived Usefulness (PU) and Perceived Ease of Use (PEOU). While TAM addresses the cognitive aspects of technology adoption, trust extends it by capturing emotional and psychological dimensions \cite{kohn_measurement_2021}, making it a critical factor in overcoming situations involving uncertainty, dependency, and risks and enhancing acceptance \cite{gefen_trust_2003, vorm_integrating_2022, ghazizadeh_extending_2012, lotfalian_saremi_trust_2024}.

Trust has been defined in various ways by scholars \cite{lee_trust_2004, mayer_integrative_1995, muir_trust_1987}, with a widely accepted definition describing it as an individual's willingness to take risks and delegate critical tasks and responsibilities to another party, such as a technological system \cite{lee_trust_2004}. In healthcare, however, clinicians often hesitate to rely on automated systems, choosing instead to depend on their own expertise. This reluctance creates significant barriers to the acceptance and adoption of advanced technologies. Considering trust as a critical factor in technology acceptance \cite{choudhury_effect_2022, glikson_human_2020, tucci_factors_2022}, many studies have focused on understanding how it can be built between users and technological systems.  

Some frameworks, such as Muir’s trust model \cite{muir_trust_1987} and Mayer et al.'s influential model focusing on ability, benevolence, and integrity \cite{mayer_integrative_1995}, provide foundational insights into trust. In the context of human-technology interactions, Parasuraman et al. \cite{parasuraman_humans_1997} expanded the understanding of trust by examining user reliance on automation and its potential for misuse or neglect. They highlighted the risks of "automation abuse," where over-automation can undermine human oversight, leading  to errors or reduced performance. Studies such as \cite{asan_artificial_2020} highlight the risks associated with overreliance on automation, including reduced monitoring of systems and the potential for biased decisions. These findings highlight the importance of designing systems that emphasize on calibrated trust to improve the humans' task performance \cite{asan_artificial_2020, lee_trust_2004, vereschak_how_2021, naiseh_how_2023, naiseh_explainable_2021, snijders_humans_2023}.

Trust formation in human-technology interactions has been widely explored, with Hoff and Bashir \cite{hoff_trust_2015} categorizing trust development into dispositional trust, situational trust, and learned trust. Building on these categories, other researchers have proposed alternative frameworks. For instance, Jermutus et al. \cite{jermutus_influences_2022} classify trust factors into user-related, AI-related, and contextual factors, while Lotfalian et al. \cite{lotfalian_saremi_survey_2021} identify three key categories: human user attributes, intelligent system design, and task characteristics. Schaefer et al. \cite{schaefer_meta-analysis_2016} mentioned the factors differently as automated partner, the environment in which the task is occurring, and
characteristics of the human interaction partner.

AI-related factors, such as explainability \cite{leichtmann_explainable_2024}, transparency \cite{lotfalian_saremi_trust_2024}, and complexity \cite{lehmann_risk_2022}, are fundamentals to enhancing user trust and reliance on AI systems. Improvements in these areas can directly address user uncertainties, ensuring outputs are clear, interpretable, and reliable. By enhancing these capabilities, AI systems empower users to make informed decisions with greater confidence, supporting appropriate reliance and facilitating broader system adoption.
Explainability has been recognized as a key factor in improving trust and reliance on AI systems \cite{tucci_factors_2022, shin_effects_2021, bernardo_affective_2023, naiseh_explainable_2021,cecil_explainability_2024, leichtmann_explainable_2024}. Users need to understand how the system generates its decisions and the reasoning behind the algorithms \cite{jermutus_influences_2022}.

\subsection{AI Explainability}
Explainable Artificial Intelligence (XAI) has become essential in fields like medicine\cite{holzinger_causability_2019, naiseh_explainable_2021}, finance, and autonomous systems, where understanding AI decisions is critical. As complex ``black-box'' models like deep learning grow in use, their lack of transparency raises concerns about trust and accountability. XAI aims to address this by providing clear, human-understandable explanations, making AI systems more transparent and reliable \cite{wang_rationality_2024}. This not only builds user confidence but also supports effective human-AI collaboration \cite{wang_rationality_2024, zhang_effect_2020, leichtmann_explainable_2024, jermutus_influences_2022}.

In the literature, various explainability methods have been introduced to address the black-box nature of AI models, which are often categorized into local and global approaches. Local methods, such as LIME (Local Interpretable Model-agnostic Explanations) \cite{ribeiro__2016}, Saliency Maps \cite{simonyan_deep_2013}, Grad-CAM (Gradient-weighted Class Activation Mapping) \cite{selvaraju_grad-cam_2020}, Integrated Gradients \cite{sundararajan_axiomatic_2017},and example-based methods, such as counterfactual explanations \cite{wachter_counterfactual_2017} and prototype-based techniques \cite{kim_examples_2016}, focus on explaining individual predictions by identifying specific features that influenced a particular output. In contrast, global methods, like SHAP (SHapley Additive Explanations) \cite{lundberg_unified_2017}, Feature Importance, and Partial Dependence Plots (PDPs) \cite{friedman_greedy_2001}, provide insights into the overall behavior of the model, showing how features contribute to predictions across the entire dataset.

Several studies in the literature have explored the impact of explainability methods on participants' behavior, particularly focusing on trust and adoption. Some studies have demonstrated positive effects, showing that explainability enhances trust and supports adoption.  For example, Wang et al. \cite{wang_rationality_2024} found that using SHAP improved decision accuracy and behavioral trust in sales prediction tasks. similarly, Leichtmann et al. \cite{leichtmann_effects_2023} conducted a 2×2 between-subject study on a mushroom-picking task, using Grad-CAM as an attribution-based explanation and ExMatchina as an example-based explanation. Both methods were found to significantly improve trust in AI systems, highlighting their effectiveness in enhancing user confidence. 

However, other studies have reported limited or no significant effects of explainability on trust. For instance, in our previous study \cite{rezaeian_impact_2024}, no improvement was observed when providing different levels of explainability, such as AI confidence scores, localization, and other methods. Ahn et al. \cite{ahn_impact_2024} found that while SHAP and LIME increased interpretability, they did not significantly enhance trust or task performance, emphasizing the role of outcome feedback over explainability. Similarly, Cecil \cite{cecil_explainability_2024} observed that explainability methods such as saliency maps and visual charts had minimal impact on mitigating the negative effects of incorrect AI advice in human resource management. Zhang et al. \cite{zhang_effect_2020}, also,  used SHAP as a local explainability method in an income prediction task but found no significant impact on trust or user behavior. 

Additionally, research comparing the effectiveness of different XAI methods has shown mixed results, with some methods being more impactful than others. For instance, Alam et al. \cite{alam_examining_2021} reported that richer explanations, such as a combination of written, visual, and example-based methods, were more effective in improving satisfaction and trust in AI diagnosis systems compared to simpler approaches. Similarly, Cai et al. \cite{cai_effects_2019} explored both normative and comparative example-based explanations in a sketch recognition task. Their findings revealed that normative explanations positively impacted system comprehension and user trust, while comparative explanations did not show significant effects. 
Wang et al. \cite{wang_are_2021} also evaluated XAI methods in recidivism and forest cover prediction tasks, comparing global approaches like feature importance with local methods such as counterfactual explanations and nearest-neighbor examples. The study found that the effectiveness of XAI largely depends on users' domain expertise, providing insights for tailoring XAI to support decision-making. The study most closely aligned with the goals of our research is by Evans et al. \cite{evans_explainability_2022}, which investigates how digital pathologists interact with various XAI tools, including saliency maps, concept attributions, prototypes, trust scores, and counterfactuals, to enhance clarity in medical imaging recommendations. While their study provides valuable insights into pathologists’ preferences for AI guidance, it does not examine how these tools and varying levels of explainability influence human-related behaviors, such as trust, reliance, cognitive load, and decision-making, which is a central focus of our work. 

Several studies have investigated AI confidence scores as an explainability feature to build trust. Zhang et al. \cite{zhang_effect_2020} found that confidence scores help calibrate trust by providing insights into the certainty of predictions. McGuirl and Sarter \cite{mcguirl_supporting_2006} showed that continually updated confidence information improves trust calibration in automated decision support systems. Similarly, Antifakos et al. \cite{antifakos_towards_2005} emphasized the importance of how confidence is presented in an automatic notification device. Rechkemmer and Yin \cite{rechkemmer_when_2022} further demonstrated that confidence scores significantly influence trust, although accuracy remains a stronger factor.

\subsection{Cognitive Load}

Cognitive load refers to the mental effort required to process information and complete tasks \cite{choudhury_impact_2023}. Cognitive Overload can negatively impact decision-making and reduce trust in automated systems, as users may feel overwhelmed or struggle to understand the system's outputs \cite{sweller_cognitive_1998}..

Research has explored the relationship between cognitive load and trust across various domains. For instance, Choudhury et al. \cite{choudhury_impact_2023} found that increased cognitive workload reduced the intent to use a blood utilization calculator (BUC), an AI-based decision support system in healthcare. This underscores the importance of designing AI systems that minimize cognitive load to foster trust and encourage adoption. However, not all studies find a direct link between explainability and cognitive load; for example, Wang et al. \cite{wang_using_2024} reported no significant differences in cognitive load between groups provided with XAI explanations versus AI-only systems.

In the context of explainable AI, the effect of explainability on cognitive load varies depending on the context and implementation. Few studies provide evidence of this variability. For instance, Herm \cite{herm_impact_2023} conducted an empirical study in medicine, using chest X-ray images of COVID-19-infected patients, and demonstrated how different XAI methods influence cognitive load. Similarly, Kaufman et al. \cite{kaufman_effects_2024} explored cognitive load in the context of autonomous driving, and Hudon et al. \cite{hudon_explainable_2021} examined the effects of XAI in image classification tasks. These findings highlight the critical need to consider human cognitive factors when designing XAI systems to ensure explanations are intuitive and do not inadvertently increase cognitive load. By carefully balancing transparency and usability, XAI can enhance trust without overwhelming users.

\section{Method}
\subsection{Overview}
This experimental study addresses research questions using online experiments with multiple intervention conditions. Each condition was designed to include different levels of explainability, to evaluate clinicians' attitudes toward an AI-based Clinical Decision Support System (CDSS) for breast cancer diagnosis. During the experiment, participants interacted with the CDSS across various diagnostic cases, assessing their trust and agreement with the system's suggestions. In addition, participants completed surveys provided demographic information before the experiment and evaluated their mental demands and stress after using each CDSS variant.

The following sections provide a detailed description of the study participants, experimental design and procedures, the specific AI configurations employed across intervention conditions, measurements, and the analytical methods used.

\subsection{Participants}
A total of 28 participants were recruited between January and August 2024 through medical associations, social platforms, and professional networks. All participants were U.S.-based, fluent in English, and over the age of 18. They were compensated \$80 for their participation. The sample comprised primarily radiologists (\(\approx60\%\)), with the remaining participants identified as oncologists (\(\approx18\%\)) or other healthcare professionals involved in breast cancer care (\(\approx22\%\)).

The demographic characteristics of the participants were as follows: The majority were male (\(\approx60\%\)), and the mean age was 42.6 years (\(SD = 12.1\)), with most participants (\(\approx53\%\)) in the 35-45 age. In terms of race and ethnicity, the majority identified as White (\(\approx68\%\)), followed by Asian (\(\approx14\%\)), Middle Eastern or North African (\(\approx7\%\)), Hispanic or Latino (\(\approx3\%\)), and other backgrounds (\(\approx7\%\)).

\subsection{Experiment Design and Procedure}

The primary goal of this study was to evaluate the effect of different levels of explainability on participants' cognitive load. We used an interrupted time series design \cite{hartmann_interrupted_1980}, where participants sequentially interacted with a breast cancer Clinical Decision Support System (CDSS) across multiple interventions, each offering a distinct level of explainability. Participants completed these interventions in a particular order, with key variables measured at multiple points to capture changes over time.

The main task required participants to diagnose ultrasound images of breast tissues as either healthy, benign, or malignant. Initially, they made these diagnoses without any decision support. As they progressed through the interventions, they were gradually provided with diagnostic suggestions from the CDSS, with varying levels of explanation generated through machine learning techniques. The experiment included pre-experiment surveys to collect demographic information and baseline attitudes toward AI. Each intervention involved diagnosing 10 images, followed by a brief post-intervention survey.

The entire experiment was conducted online on a web application developed using the Python-based Dash framework. The online application platform integrated all stages, from signing the consent form and completing surveys to participating in the experiment sessions and collecting data. This setup allowed us to recruit participants nationwide without requiring in-person supervision.

After recruitment, participants received a video tutorial and an experiment link via email. This approach ensured that only those genuinely interested and informed about the study participated, helping to maintain data authenticity. Upon accessing the link, participants first signed a consent form, completed a pre-experiment survey, and then proceeded through a series of experiment sessions as follows:

\begin{figure*}[!h]
    \centering
    \includegraphics[width=1\linewidth]{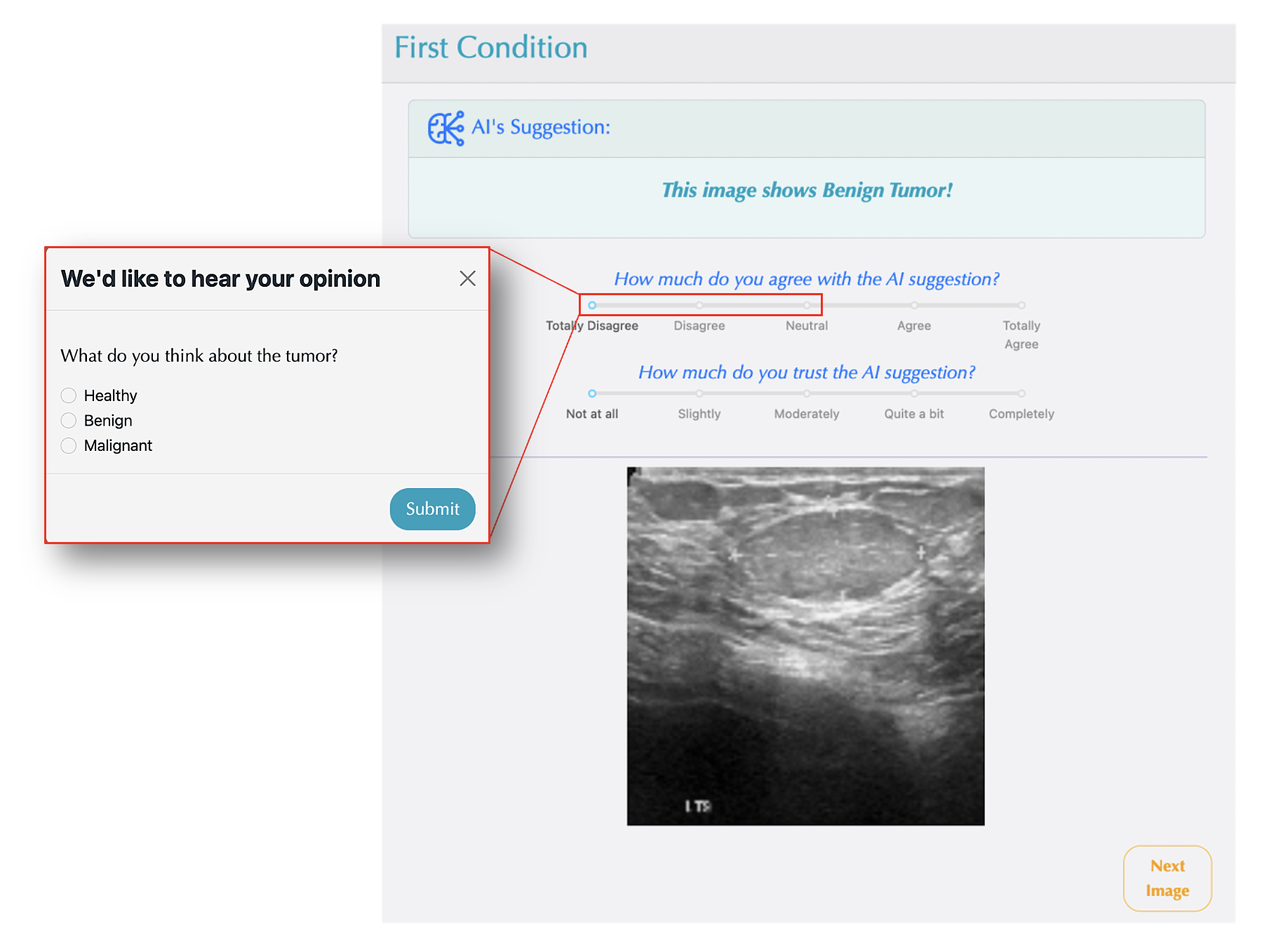}
    \caption{Experiment web interface with an interactive assessment pop-up window, where participants' opinions are recorded when their diagnostic judgments differ from those of the AI.}
    \label{fig:Main-platform}
    \vspace{2em}

\end{figure*}

\begin{itemize}
    \item \textbf{\textit{Baseline (Stand-alone):}} Clinicians review breast cancer tissue images and make diagnostic decisions independently, without any suggestions.
    \item \textbf{\textit{Intervention I (Classification):}} Clinicians receive diagnostic recommendations (healthy, benign tumor, malignant tumor) from the system, but without any explanatory details.
    \item \textbf{\textit{Intervention II (Probability Distribution):}} Expanding on the previous intervention, this stage includes probability scores for each diagnostic option (healthy, benign tumor, malignant tumor).
    \item \textbf{\textit{Intervention III (Tumor Localization):}} In addition to probability estimates, the system highlights the suspected location of the tumor within the breast tissue images.
    \item \textbf{\textit{Intervention IV (Enhanced Localization with Confidence Levels):}} Building on the third intervention, clinicians are provided with tumor localization information that includes areas marked with both high and low confidence levels.
\end{itemize}

In the baseline condition, participants analyzed ultrasound images independently and provided their diagnosis as healthy, benign, or malignant. In the subsequent conditions, participants were given diagnostic suggestions and asked to rate their agreement and trust in these suggestions on a 5-point Likert scale (as shown in \textit{\Cref{fig:Main-platform}}). If their agreement rating was below 3, they were prompted to make their own diagnosis and share their opinion on the image (see \textit{\Cref{fig:Main-platform}}).

\begin{figure*}
    \centering
    \includegraphics[width=0.75\linewidth]{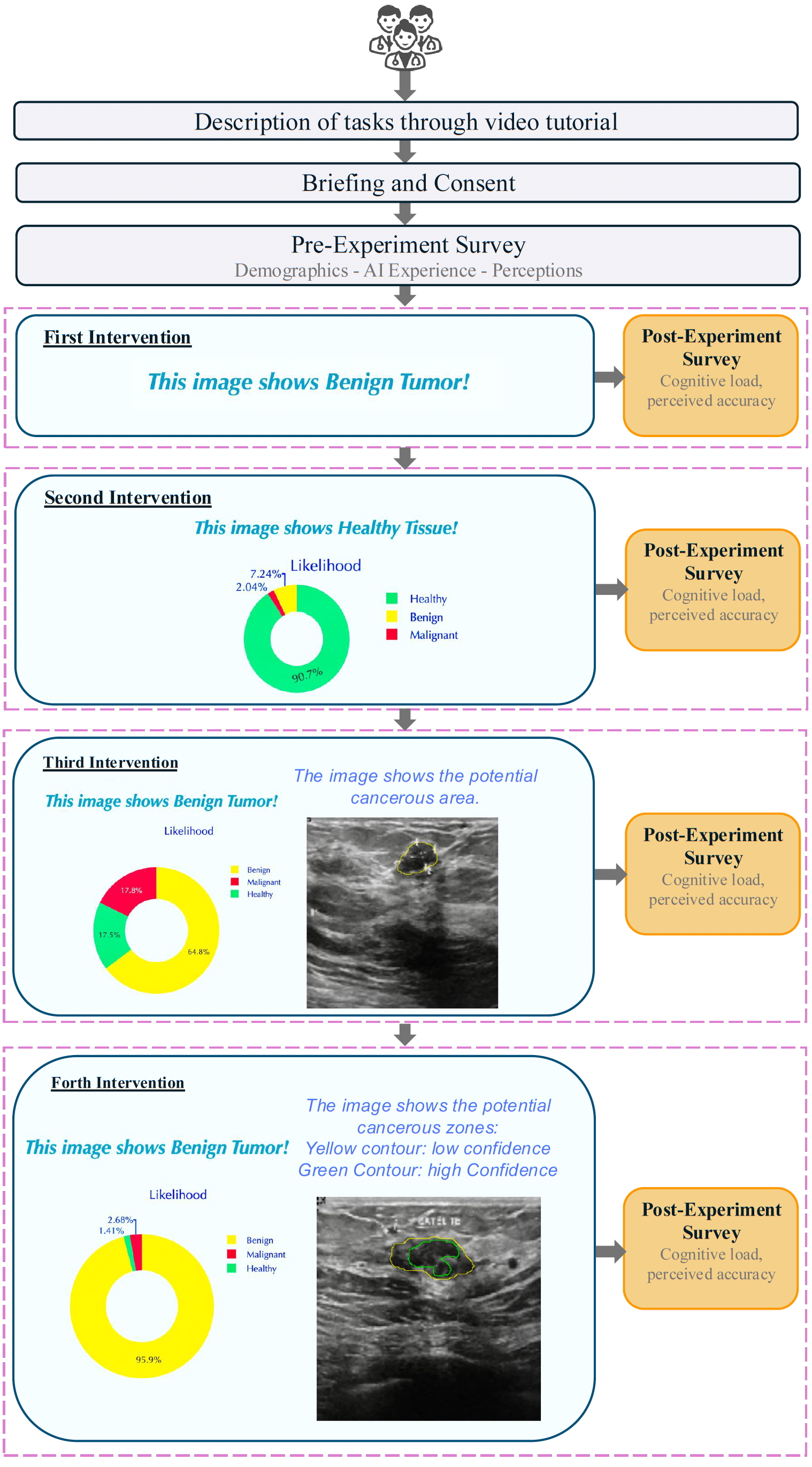}
    \caption{Experiment Procedure}
    \label{fig:Procedure}
\end{figure*}

After completing each intervention, participants proceeded to a post-intervention survey designed to capture specific insights about that stage. Using a 5-point Likert scale, they rated aspects such as cognitive load and other relevant factors. The full procedure for the experiment is outlined in \textit{\Cref{fig:Procedure}}.

The AI system supporting the CDSS is based on our previous work~\cite{rezaeian_architecture_2024} and combines a U-Net model for precise image segmentation with a Convolutional Neural Network (CNN) for diagnostic classification. The U-Net architecture \cite{ronneberger_u-net_2015} is specifically designed to delineate the boundaries of suspicious regions within breast ultrasound images, effectively highlighting areas that may contain cancerous tissue. This segmentation step assists in localizing potential tumors and provides visual cues that clinicians can use to focus their analysis in 3rd and 4th Interventions.
Following this step, the segmented images are analyzed by the CNN, a widely used method for analyzing medical images \cite{abdelrahman_convolutional_2021, khazrak_addressing_2024, yadav_deep_2019}, which categorizes each image into one of three diagnostic classes: healthy, benign, or malignant. This classification model was trained using a publicly available breast cancer ultrasound dataset containing 780 images \cite{al-dhabyani_dataset_2020}. The training process enabled the model to learn and recognize distinguishing features associated with each diagnostic category, achieving an overall diagnostic accuracy of 81\%.

\subsection{Assessment Measures}
In the pre-experiment survey, we assessed participants' perceptions of AI through three key measures: AI role, AI usefulness, and complexity perception.

\subsubsection*{AI Role}
To understand participants' perspectives on AI’s potential in healthcare, we asked them to indicate their level of agreement with the statement, ``Artificial Intelligence will play an important role in the future of medicine.'' Responses were captured on a 5-point Likert scale, from Totally Disagree to Totally Agree.
\subsubsection*{AI Usefulness}
This measure examined how relevant participants perceived AI to be for their specific roles. Participants rated their agreement with the statement, ``AI would be useful in my job.'' providing insight into AI's perceived practical benefits in their professional contexts.
\subsubsection*{Complexity Perception}
To assess perceived barriers to AI adoption, participants responded to the statement, ``There are too many complexities and barriers in medicine for AI to help in clinical settings.'' This measure, also rated on a 5-point Likert scale, helped identify potential challenges related to AI integration in clinical practice.\\

During the experiment, we recorded participants' responses and interactions with the AI system using the following key measures:
\subsubsection*{Trust}
We measured participants' trust in AI on a 5-point Likert scale from 0 (No trust) to 5 (Complete trust). For each image, they rated their trust in the AI's suggestions, answering, ``How much do you trust the AI system?''
\subsubsection*{Agreement}
We measured participants' agreement in AI's suggestions on a 5-point Likert scale from 0 (No trust) to 5 (Complete trust). For each image, they rated their agreement in the AI's suggestions, answering, ``How much do you agree with AI suggestion?''
\subsubsection*{Performance}
Participants' performance was evaluated based on their decisions for each image during the experiment. Performance was assessed in two ways: If a participant's agreement level with the AI suggestion was above neutral, their decision was considered aligned with the AI’s suggestion. When the agreement level was below neutral, participants independently selected a diagnosis of “Healthy,” “Benign,” or “Malignant”. Performance for each intervention was calculated by comparing their choices to the ground truth across ten images, expressed as a percentage to represent decision accuracy relative to correct diagnoses.

\subsubsection*{Diagnosis Duration}
We recorded the time participants took to make decisions for each image during the experiment.
\subsubsection*{AI Confidence Score}

This measure represents the classification probabilities by the AI model in each category. Starting from 2nd intervention, we displayed the probability of classification in percentage for all three categories for each image. We recorded the highest probability assigned by the AI for each prediction. For images shown in 1st intervention, we did not record a value for this variable. For subsequent interventions, we categorized this variable as low confidence for predictions with less than 90\% likelihood and high confidence for those with 90\% or higher.

In the post-experiment survey after each intervention, we assessed participants' experiences through two key questions extracted from NASA Task Load Index (TLX) \cite{hart_development_1988} about mental demand and stress.
\subsubsection*{Mental Demand}
After each intervention, we assessed mental demand by asking participants, ``How mentally demanding was the task?'' Responses were given on a 5-point Likert scale, from Very Low to Very High. This measure helped us understand the cognitive load experienced by participants at each stage.
\subsubsection*{Stress}
After each intervention, we measured Stress by asking participants, ``How stressed were you?'' Responses were collected on a 5-point Likert scale, ranging from Very Low to Very High. This measure provided insights into participants' stress levels throughout the experiment.

\section{Results}
\subsection{Impact of AI Explainability on Cognitive Load: Mental Demand and Stress}
The mixed-effects linear model analyses examined the impact of different levels of AI explainability on clinicians’ mental demand and stress. The models included intervention conditions as fixed effects and participant groups as random effects, controlling for inter-individual variability.

\subsubsection*{Mental Demand}
The model for mental demand, as shown in \textit{\Cref{tab:Cognitiveload}} indicated that the intercept, representing the baseline (1st Intervention), had a significant effect ($\beta = 1.714$, $p < 0.05$), suggesting that participants experienced a moderate level of mental demand in the baseline condition. However, none of the intervention levels (2nd, 3rd, or 4th Intervention) showed a statistically significant difference in mental demand compared to the baseline, although a slight decrease was observed. Specifically, the coefficients for the 2nd, 3rd, and 4th interventions were $\beta = -0.036$, $\beta = -0.179$, and $\beta = -0.107$, respectively, all with $p > 0.05$.

\begin{table}[hb]
\vspace{1em}

\centering
\caption{Mixed linear model results: Effect of AI explainability on Cognitive Load:}
\small
\resizebox{0.95\columnwidth}{!}{%
\begin{tabular}{lcccc}
\hline
\textit{Cognitive Load}           & \multicolumn{2}{c}{\textbf{Mental Demand}}  & \multicolumn{2}{c}{\textbf{Stress}}  \\
\hline
\textbf{Variable}    & \textbf{$\beta$} & \textbf{$p$} & \textbf{$\beta$} & \textbf{$p$} \\
\hline
Intercept            &1.714      & \textbf{0.000}      &1.071      &\textbf{0.000}      \\
2nd Intervention     &-0.036     & 0.822               &0.214      &  0.185             \\
3rd Intervention     &-0.179     & 0.262               &0.393      & \textbf{0.015}     \\
4th Intervention     &-0.107     & 0.501               &0.214      & 0.185              \\
Group Var            &-0.881     &                     & 0.785     &                    \\
\hline

\end{tabular}%
}
\label{tab:Cognitiveload}
\end{table}

\subsubsection*{Stress}
For stress (\textit{\Cref{tab:Cognitiveload}}), the intercept (First Intervention) was also significant ($\beta = 1.071$, $p < 0.05$), indicating a moderate baseline level of stress. Interestingly, the 3rd Intervention (which introduced tumor localization with probability estimates) significantly increased stress compared to the baseline ($\beta = 0.393$, $p = 0.015$), suggesting that this specific level of explainability may add to participants' stress levels. The 2nd and 4th interventions, however, showed an increase in comparison to the first intervention but did not produce significant effects ($\beta = 0.214$, $p = 0.185$ for both), indicating no significant stress changes at these levels of explainability.

\subsection{Impact of AI Confidence Score on Users' Behavior}

The mixed-effects model was used to assess the influence of AI confidence score (No confidence, low and high) on four key outcomes: trust, agreement, performance, and diagnosis duration. The results are summarized in Table \ref{tab:Confidence}.

\subsubsection*{Trust}
Low confidence score significantly decreased trust ($\beta = -0.163$, $p = 0.023$), indicating that participants reported lower trust when the AI’s confidence in its predictions was low compared to when no confidence score was provided. High confidence, while not significantly different from the baseline ($\beta = 0.103$, $p = 0.169$), showed a slight increase in trust.
\subsubsection*{Agreement}
Agreement levels also showed a significant decrease with low AI confidence score ($\beta = -0.186$, $p = 0.005$), indicating that participants were less likely to agree with the AI’s recommendations when the confidence score was low. Similar to trust, high confidence score did not have a significant effect on agreement ($\beta = 0.108$, $p = 0.121$).

\begin{table*}
\centering
\caption{Mixed linear model results: Effect of AI confidence score on trust, agreement, performance, and diagnosis duration}
\normalsize
\resizebox{0.8\textwidth}{!}{
\begin{tabular}{lcccccccc}
\hline
\textit{AI Confidence Score}           & \multicolumn{2}{c}{\textbf{Trust}}  & \multicolumn{2}{c}{\textbf{Agreement}}  & \multicolumn{2}{c}{\textbf{Performance}}  & \multicolumn{2}{c}{\textbf{Diagnosis Duration}}  \\
\hline
\textbf{Variable}    & \textbf{$\beta$} & \textbf{$p$} & \textbf{$\beta$} & \textbf{$p$} & \textbf{$\beta$} & \textbf{$p$} & \textbf{$\beta$} & \textbf{$p$} \\
\hline
Intercept            & 3.031      & \textbf{0.000}   & 3.224      & \textbf{0.000}  & 0.738     &\textbf{0.000} &0.260  &\textbf{0.000} \\
Low Confidence       & -0.163     & \textbf{0.023}   & -0.186     & \textbf{0.005}  & -0.087    &0.675          &0.131 &\textbf{0.009}\\
High Confidence      & 0.103      & 0.169            & 0.108      & 0.121           & -0.015    &\textbf{0.020} &-0.012  & 0.818 \\
Group Var            & 0.292      &                  & 0.238      &                 & 0.005     &               &0.094  &\\
\hline
\end{tabular}%
}
\label{tab:Confidence}
\vspace{-1em}
\end{table*}

\subsubsection*{Performance}

Performance, as an outcome measure, showed a unique pattern. High confidence was associated with a small but statistically significant decrease in performance ($\beta = -0.015$, $p = 0.020$), suggesting that participants may have been overly reliant on the AI’s suggestions, whose accuracy was approximately 80\%, potentially reducing their diagnostic accuracy. In contrast, low confidence did not significantly affect performance ($\beta = -0.087$, $p = 0.675$).

\subsubsection*{Diagnosis Duration}
For diagnosis duration, low confidence led to a small but significant increase in the time taken ($\beta = 0.131$, $p = 0.009$), suggesting that participants may have spent more time on their diagnoses when the AI expressed lower confidence. High confidence, however, did not significantly impact diagnosis duration compared to the baseline ($\beta = -0.012$, $p = 0.818$).\\

Overall, low confidence from the AI system significantly decreased trust and agreement while increasing diagnosis duration, suggesting that participants were more cautious and took additional time on diagnoses when the AI’s confidence was low. In contrast, high confidence did not produce significant changes in these measures, highlighting a potential asymmetry in how confidence scores affect participant behavior. These findings emphasize the importance of AI confidence displays to encourage balanced reliance and avoid overconfidence in the AI's recommendations.

\subsection{Assessing the Effect of Participant Demographics on Study Outcomes}

The ANOVA results presented in \textit{\Cref{tab:ANOVA}} represent the influence of demographic variables on various outcome measures concerning perceptions and experiences of participants. The outcome measures which were considered in this regard are mental demand, stress, AI role, complexity perception, and AI usefulness. Significant effects are identified based on the F-statistic along with p-values across these variables.

\begin{table*}[]
\vspace{1.5em}

\centering
\caption{ANOVA Test Results: Effect of Categorical Variables on Outcome Measures}
\resizebox{0.98\textwidth}{!}{
\normalsize
\begin{tabular}{lcccccccccccc}
\hline
                     & \multicolumn{2}{c}{\textbf{Mental Demand}}  & \multicolumn{2}{c}{\textbf{Stress}} & \multicolumn{2}{c}{\textbf{AI Role}} & \multicolumn{2}{c}{\textbf{Complexity Perception}}  & \multicolumn{2}{c}{\textbf{AI Usefulness}}\\
\hline
\textbf{Variable}    & \textbf{F} & \textbf{p} & \textbf{F} & \textbf{p} & \textbf{F} & \textbf{p} & \textbf{F} & \textbf{p} & \textbf{F} & \textbf{p} \\
\hline
Age          & 35.91      & \textbf{8.64e-16} & 12.47 & \textbf{5.51e-07} & 10.02 & \textbf{8.00e-06} & 62.31 & \textbf{1.03e-22} & 14.23 & \textbf{8.78e-08} \\
Experience Bracket   & 8.91       & \textbf{2.77e-04} & 1.16  & 0.317            & 15.75 & \textbf{1.00e-06} & 1.46  & 0.237             & 5.91  & \textbf{3.76e-03} \\
Job                  & 1.93       & 0.151            & 1.11  & 0.334            & 8.26  & \textbf{4.82e-04} & 11.78 & \textbf{2.58e-05} & 4.48  & \textbf{1.37e-02} \\
Race                 & 2.82       & \textbf{2.88e-02} & 2.51  & \textbf{0.047}            & 8.49  & \textbf{6.00e-06} & 4.78  & \textbf{1.44e-03} & 12.24 & \textbf{3.99e-08} \\
Gender               & 1.46       & 0.231            & 0.015 & 0.901            & 0.31  & 0.579            & 92.97 & \textbf{6.61e-16} & 23.68 & \textbf{4.30e-06} \\
\hline
\end{tabular}
}
\label{tab:ANOVA}
\end{table*}

Results reveal that \textbf{age} is a highly significant factor across all outcome measures, with extremely low p-values (all below $10^{-5}$), indicating strong relationships between age groups and their responses. Similarly, \textbf{experience bracket} significantly affects certain outcome measures, namely, \textit{mental demand} ($p = 2.77 \times 10^{-4}$), \textit{AI role} ($p = 1.00 \times 10^{-6}$), and \textit{AI usefulness} ($p = 3.76 \times 10^{-3}$), pointing to a relationship between participants’ experience levels and their perceptions of these specific factors.

\textbf{Job} role also shows a significant effect on perceptions, especially regarding \textit{AI role} ($p = 4.82 \times 10^{-4}$), \textit{complexity perception} ($p = 2.58 \times 10^{-5}$), and \textit{AI usefulness} ($p = 1.37 \times 10^{-2}$). This implies that job responsibilities may shape how participants view AI functionality and complexity. \textbf{Race} has a significant impact on all outcome measures as well, highlighting potential differences in AI-related perceptions across racial backgrounds. Lastly, \textbf{gender} plays a significant role specifically in perceptions of \textit{complexity perception} ($p = 6.61 \times 10^{-16}$) and \textit{AI usefulness} ($p = 4.30 \times 10^{-6}$), suggesting that men and women might perceive AI complexity and usefulness differently.

When it comes to AI usefulness, female oncologists perceive AI as the most useful, whereas male radiologists show lower ratings, suggesting that gender and job role impact how useful AI is perceived within these fields. Finally, for AI role, radiologists report a generally high perception of AI’s role, with male oncologists rating it lowest, indicating that radiologists may view AI as more integral to their field compared to other job roles. These findings underscore how gender and professional role shape perceptions of stress, mental demand, task complexity, and the role and usefulness of AI, with notable differences seen in the oncologist and radiologist groups.

\begin{figure}
\vspace{1.5em}

    \centering
    \includegraphics[width=1\linewidth]{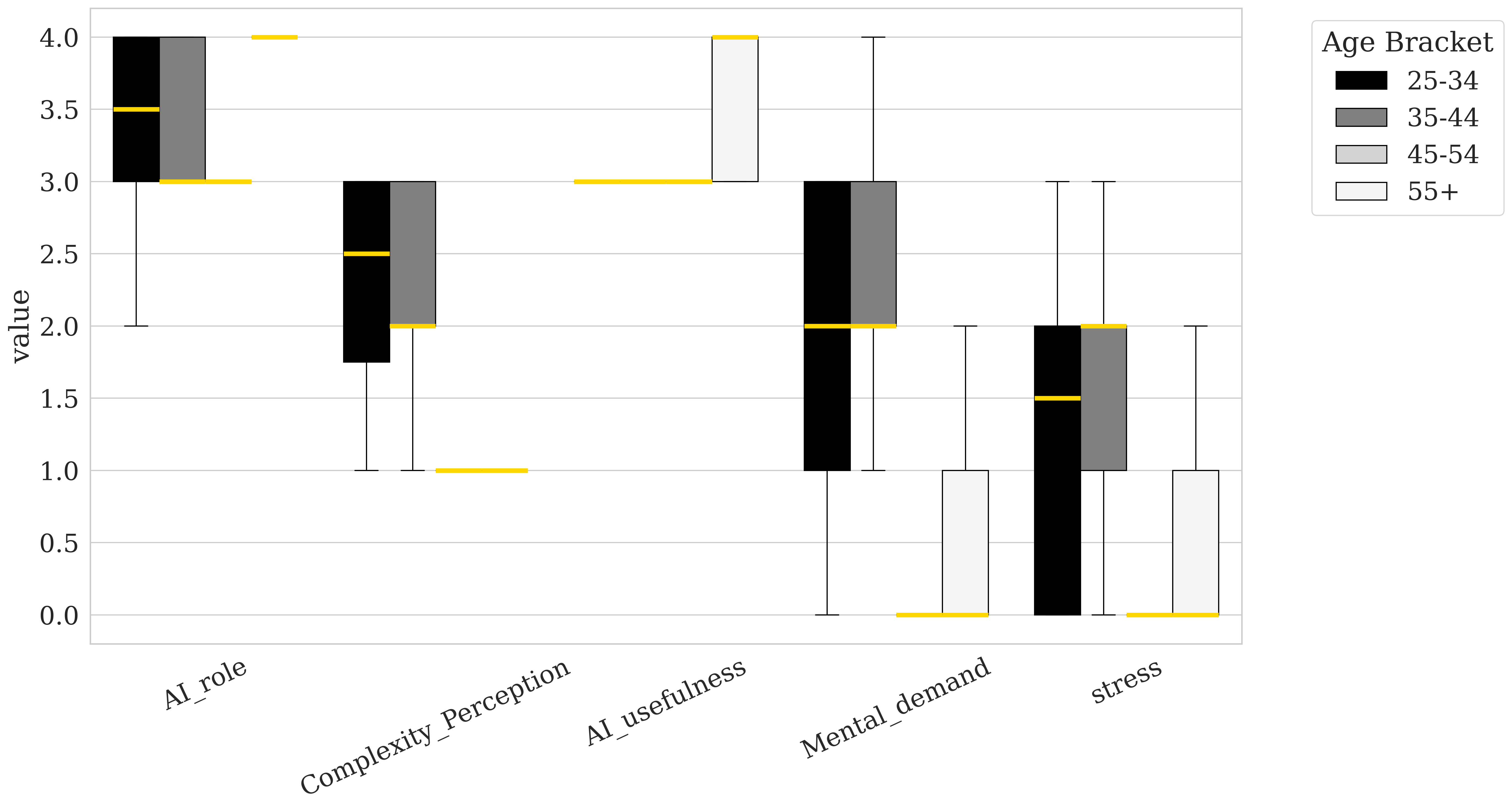}
        \caption{Distribution of AI-related measures by age.}
    \label{fig:age}
\end{figure}

The box plots illustrate how participants' perceptions of AI role, complexity perception, AI usefulness, mental demand, and stress vary across different demographic categories: age, gender, and race. For age (\textit{\Cref{fig:age}}), younger participants (25-34) tend to rate AI role and AI usefulness lower, while older participants, especially those 55 and above, rate these aspects higher, suggesting that older participants see more value in AI's role and utility. Notably, older participants report lower mental memand and stress levels, suggesting that their greater experience may reduce the cognitive load required for these tasks compared to younger participants.

\begin{figure}
    \centering
    \includegraphics[width=1\linewidth]{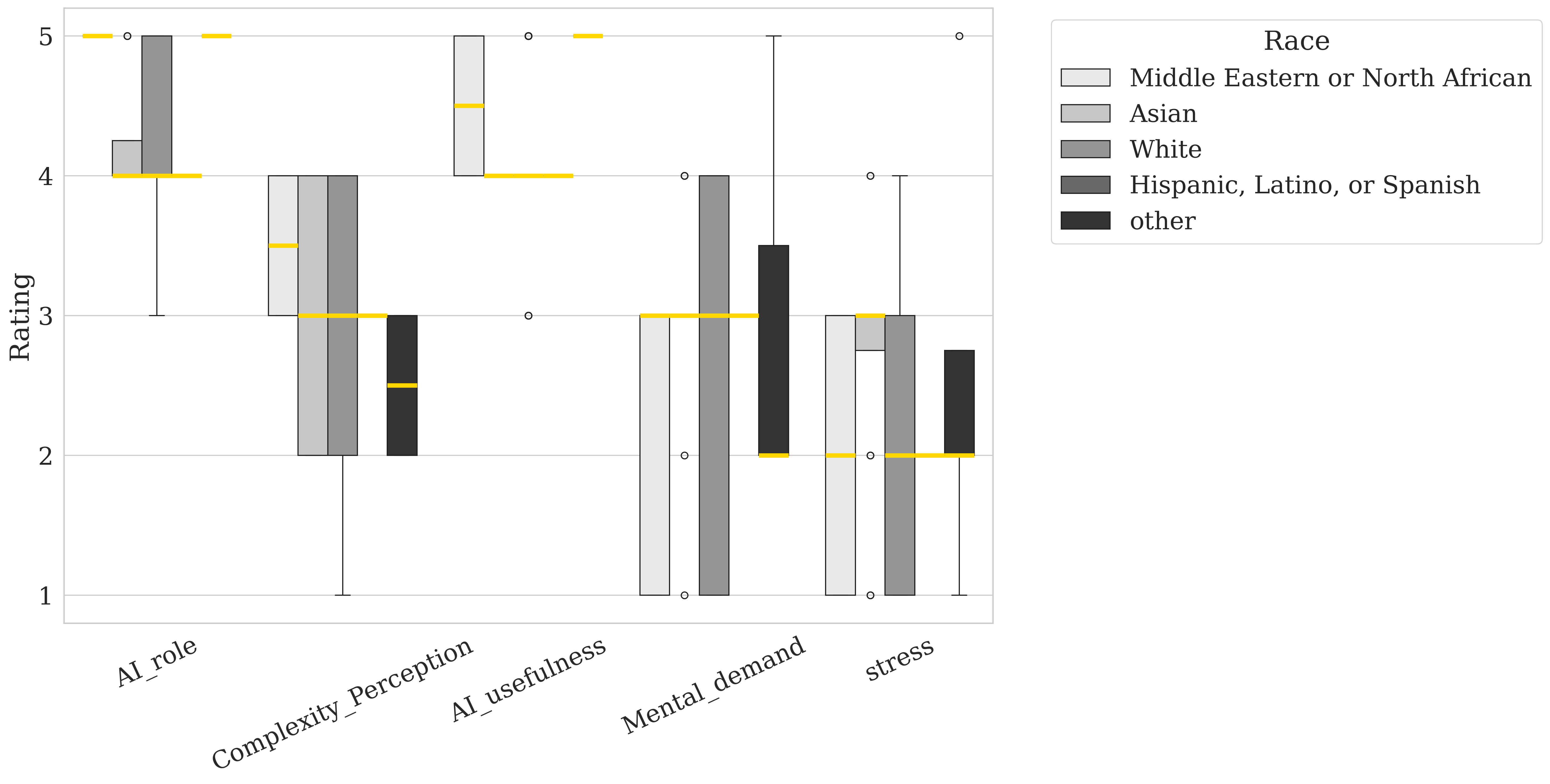}
        \caption{Distribution of AI-related measures by race.}
    \label{fig:race}
\end{figure}

\begin{figure}
    \centering
    \includegraphics[width=1\linewidth]{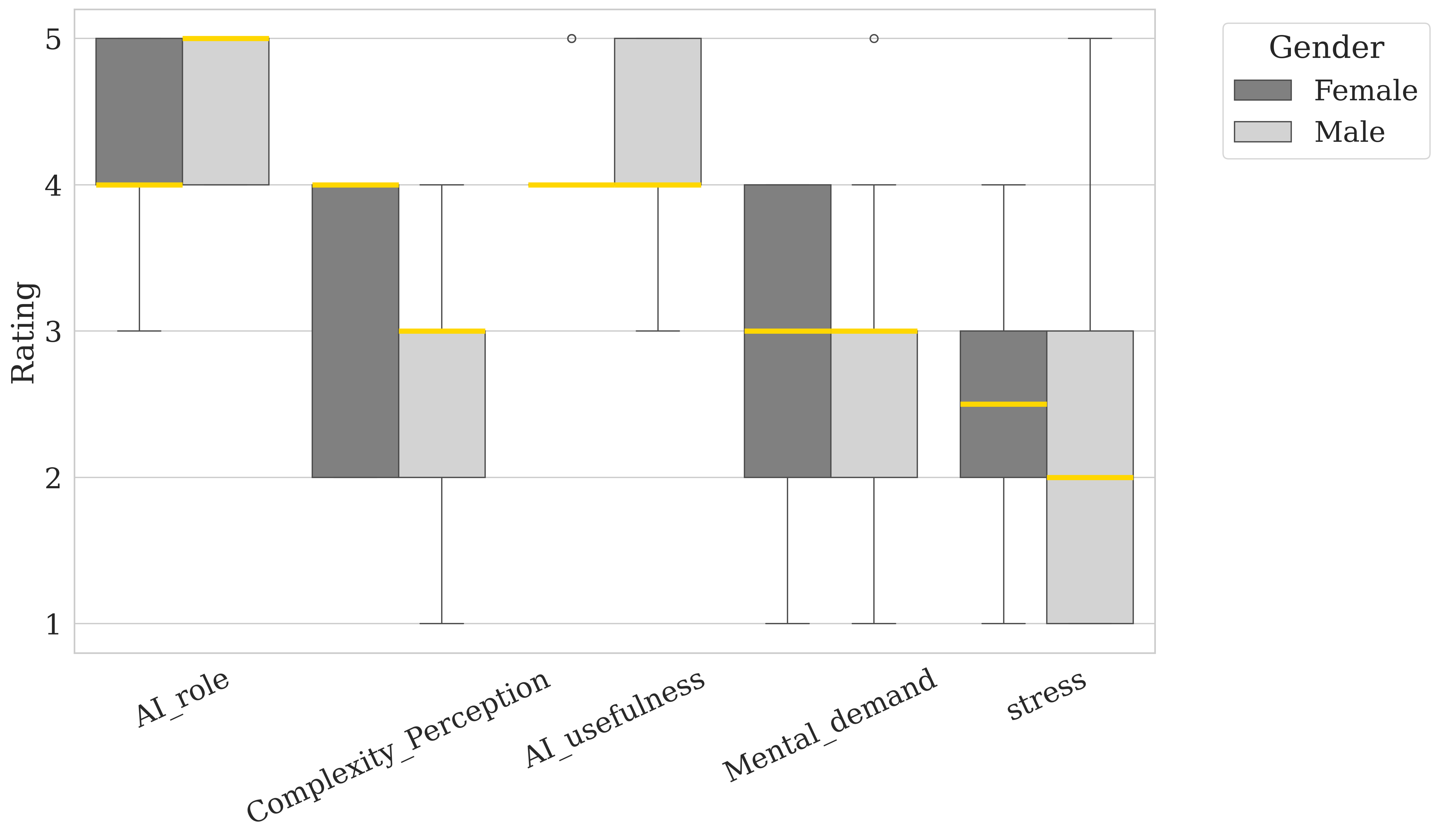}
        \caption{Distribution of AI-related measures by Gender.}
    \label{fig:gender}
\end{figure}

Gender differences (\textit{\Cref{fig:gender}}) reveal that female participants perceive the AI role as less significant compared to male participants but report higher levels of complexity and stress. This suggests that, while females may view AI as having a lesser role, they find the tasks associated with it more complex and demanding than their male counterparts. Racial differences are also notable as shown in \textit{\Cref{fig:race}}: middle eastern or north African participants rate AI role, complexity, and AI usefulness highly, indicating a generally positive perception of AI's potential and relevance. In contrast, participants from the Asian group report higher levels of Stress, suggesting that they may experience more cognitive or emotional strain when engaging with AI-related tasks. These findings suggest that age, gender, and race shape individuals’ views on AI’s utility and complexity, as well as the cognitive and emotional demands associated with interacting with AI systems.


\section{Discussion}
\subsection{Impact of AI Explainability on Cognitive Load}
Most studies focus on the effect of explainability on cognitive load by examining individual explanation models \cite{bucinca_proxy_2020, hudon_explainable_2021} or comparing different models \cite{herm_impact_2023, bucinca_trust_2021}. To answer RQ1, this study broadens the scope by exploring the impact of varying levels of AI explainability, from no explanation to increasingly detailed explanations, on cognitive load, specifically mental demand and stress.

Karran et al. \cite{karran_designing_2022} argue that any form of XAI explanation can reduce cognitive effort by aligning users' mental models with the task. Our findings align with this perspective, showing that increasing levels of explainability generally decreased mental demand. However, this reduction was not statistically significant, consistent with Wang et al. \cite{wang_using_2024}, who reported no significant differences in cognitive load between groups provided with XAI explanations and those using AI-only systems. Notably, explanations that were concise, visually clear, and focused on relevant information showed the greatest potential for reducing mental load.

In contrast, the fourth condition, which introduced more complex and detailed information, resulted in increased mental demand. This highlights that overly detailed or poorly aligned explanations can conflict with users’ mental models, requiring additional cognitive effort to process. These findings underscore the importance of designing explanations that are both clear and relevant, as their quality significantly affects cognitive load \cite{herm_impact_2023}. By prioritizing user-centered explanation design, XAI systems can better support users without introducing unnecessary cognitive strain.

On the other hand, stress levels increased for all levels of explainability, with the third condition,  tumor localization with probability estimates, causing a statistically significant rise. This increase can be attributed to the distinct nature of the information introduced in this condition, which required greater cognitive effort to interpret and apply. Consistent with previous studies, such as \cite{bucinca_proxy_2020}, explanations can impose additional cognitive demands depending on the style and complexity of the information provided.

These findings highlight the need for clear and concise XAI design. Overly complex explanations can increase cognitive demands, while appropriate information helps reduce strain and improve usability \cite{kaufman_effects_2024}.

\subsection{Effect of AI Confidence Scores on User Behavior}

Confidence scores play a crucial role in helping users calibrate their trust in AI systems, enabling them to determine when to trust or distrust AI outputs \cite{zhang_effect_2020}. The effectiveness of displaying confidence information has been observed across diverse domains, such as air traffic control, military training, and mobile applications, highlighting its broad applicability \cite{antifakos_towards_2005}. In this study, we investigated the impact of varying levels of AI confidence on trust, agreement, performance, and diagnosis duration within the healthcare domain to address RQ2.

Research shows that dynamic confidence displays can reduce automation bias. For example, \cite{mcguirl_supporting_2006} found that pilots who had access to real-time AI confidence updates were less likely to blindly follow system recommendations. Similarly, we found that low AI confidence significantly decreased trust and agreement but increased diagnosis duration. This cautious behavior aligns with \cite{antifakos_towards_2005, zhang_effect_2020}, who noted that lower confidence scores encourage users to rely more on their own judgment, leading to more thoughtful decision-making.

On the other hand, high confidence slightly increased trust, consistent with findings by \cite{zhang_effect_2020,antifakos_towards_2005}. However, it also led to a small but significant decrease in performance, likely due to overreliance on AI recommendations. This overreliance, often referred to as automation bias, has been introduced by Parasuraman and Riley \cite{parasuraman_humans_1997}, which highlight how users tend to trust high-confidence AI outputs without critically assessing them.

These findings highlights the paramount importance of designing AI confidence displays to encourage appropriate reliance on AI systems. While high confidence can enhance trustworthiness, it must be presented in a manner that mitigates automation bias and ensures balanced decision-making, particularly in critical domains such as healthcare. 

\subsection{Demographic Factors and Perceptions of AI}
To address Research RQ3, we analyzed participant demographics and their perceptions of AI, revealing significant relationships between age, gender, job role, and user interaction with AI systems. Research indicates that individuals of different ages may adopt varied strategies when assessing the trustworthiness of automated systems \cite{hoff_trust_2015}. In contrast to findings by Jermutus et al. \cite{jermutus_influences_2022}, our results show that older participants rated the role and usefulness of AI more positively, whereas younger participants reported higher levels of mental demand and stress during interactions with AI.

Gender differences also emerged as a critical factor. Female participants perceived tasks involving AI as more complex and demanding and rated the usefulness of AI lower than their male counterparts, consistent with Tung's \cite{tung_influence_2011} findings. These variations may be influenced by differences in how males and females respond to an automated system’s communication style and interface design \cite{hoff_trust_2015, tung_influence_2011, t_nomura_prediction_2008}.

Participants’ professional and technological backgrounds played a significant role in shaping their perceptions of AI. Research shows that users with higher digital literacy and greater exposure to AI systems are generally more comfortable and confident in adopting such technologies \cite{deursen_smart_nodate}. Radiologists, for instance, viewed AI as more integral to their work compared to other professional groups, likely due to their frequent exposure to AI systems specifically designed for radiology tasks. In contrast, oncologists found tasks involving AI to be more complex and demanding. This may reflect the relative lack of AI tools tailored to their specific workflows or the need for additional training to bridge gaps in familiarity and confidence. Social influences, including peer and family attitudes toward AI, also play a role in shaping users’ willingness to engage with these systems \cite{lotfalian_saremi_trust_2024}.

\subsection{Limitations}

While this study offers valuable insights, there are some important limitations to consider. Because the experiment was conducted in a controlled setting, it doesn’t fully reflect the complexities of real-world clinical environments. Factors like emotional involvement, high-pressure decision-making, and how the system fits into existing workflows weren’t part of this study. We also relied on self-reported measures like trust and cognitive load, which may not always match what participants actually think or do. Additionally, the participant group wasn’t large or diverse enough to capture the wide range of perspectives found in real-world clinical settings.

Future research should explore how these findings apply to different environments and longer-term use. By addressing these gaps, we can better understand how to design AI systems that are effective, trustworthy, and equitable in practice.

\section{Conclusion}
Artificial Intelligence (AI) has revolutionized various fields, including healthcare, by enhancing decision-making processes and outcomes. In the context of Clinical Decision Support Systems (CDSSs), the integration of explainability features has become increasingly important to ensure user trust and engagement. However, the relationship between explainability, trust, and cognitive load remains complex and requires further exploration.

This study investigated the effects of different explainability features, such as AI confidence scores, on trust and cognitive load in a controlled experimental setting. Our findings revealed some interesting relationships: high AI confidence substantially improved trust but led to overreliance, while low confidence decreased trust and agreement but increased diagnosis duration. AI confidence score also elevated stress levels, highlighting the need to carefully design these features to minimize cognitive load and avoid unintended consequences.

Demographic factors, including age, gender, and professional roles, significantly influenced participants’ perceptions of AI, emphasizing the importance of tailoring AI systems to meet the needs of diverse user groups. These findings underscore the importance of designing explainable AI systems that balance transparency and usability, ensuring they foster trust, reduce cognitive demands, and support effective decision-making in high-stakes domains like healthcare.

\bibliographystyle{ieeetr} 
\bibliography{referencesMod.bib}

\end{document}